# From Equivalence Principles to Cosmology: Cosmic Polarization Rotation, CMB Observation, Neutrino Number Asymmetry, Lorentz Invariance and CPT

Wei-Tou Ni[*)1,2]

[1]*Center for Gravitation and Cosmology, Purple Mountain Observatory,*
*Chinese Academy of Sciences, Nanjing, 210008 China*
[2]*National Astronomical Observatories, Chinese Academy of Sciences, Beijing, 100012 China*

In this paper, we review the approach leading to cosmic polarization rotation observation and present the current status with an outlook. In the study of the relations among equivalence principles, we found that long-range pseudoscalar-photon interaction is allowed. Pseudoscalar-photon interaction would induce a rotation of linear polarization of electromagnetic wave propagating with cosmological/astrophysical distance. In 2002, DASI successfully observed the polarization of the cosmological microwave background radiation. In 2003, WMAP observed the correlation of polarization with temperature anisotropy at more than 10 σ in the cosmological microwave background. From this high polarization-temperature correlation in WMAP observation, we put a limit of 0.1 rad on the rotation of linear polarization of cosmological microwave background (CMB) propagation. Pseudoscalar-photon interaction is proportional to the gradient of the pseudoscalar field. From phenomenological point of view, this gradient could be neutrino number asymmetry current, other density current, or a constant vector. In these situations, Lorentz invariance or CPT may effectively be violated. In this paper, we review and compile various results. Better accuracy in CMB polarization observation is expected from PLANCK mission to be launched next year. A dedicated CMB polarization observer in the future would probe this fundamental issue more deeply.

## §1. Introduction and summary

Cosmology has the aim of studying our universe and its contents to understand its structure, evolution and origin. During the past decades, observations poured in and cosmology became more and more an observational science. These observations include cosmic microwave background (CMB) observations, gravitational lens observations, galaxy counting and correlation observations, distance and age observations, active galaxies and quasar observations, and many more. Astronomy and astrophysics observes celestial objects in our universe whose basic constituents are elementary particles, and particle physics studies theories and observational properties of these particles. All objects and particles evolve in the arena of spacetime, and mutually interact with one another and with spacetime gravitationally. Hence, astronomy/astrophysics, particle physics, and gravitation are three disciplines which are closely related to cosmology as depicted in Fig. 1. The study of cosmology serves as a basic arena to astronomy/astrophysics, and stimulates discoveries in particle physics and gravitation.

One example that these three disciplines and cosmology are closely related is the solar neutrino problem. The deficiency of observed solar neutrinos on Earth[1] pointed to 3 potential directions: (i) whether solar model was incorrect and needed to be improved,[2] (ii) whether there is an intermediate force so that the 'gravitational constant' changes with distance and solar mass is different;[3-5] and (iii) neutrino oscillation.[6] After the stellar nuclear evolution theory had been consolidated and the deficiency of detected solar neutrino still persisted, and after experiments on intermediate force indicated that it could not be the possible explanation of solar neutrino problem, further particle experiments on neutrinos discovered neutrino oscillations. A complete theory of neutrino oscillation is

---
[*)] E-mail: wtni@pmo.ac.cn

close at hand waiting for further experimental results on neutrinos. The correct theory of neutrino oscillation will have an impact both on particle physics and on cosmology. Another example is constraints on compactification radius of higher dimensional ADD (Arkani-Hamed, Dimopoulos and Dvali)[7] models. Inverse-square-law experiments gives upper limits of 36.6 μm and 62 μm on $R_3$ (the compactification radius for 3 extra dimensions) and $R_4$ (the compactification radius for 4 extra dimensions) respectively,[8] while spectroscopic measurement on the 1s-2s muonium transition gives upper limits of 10 μm and 8.2 μm correspondingly.[9-11] The most stringent limits are 4.2 pm and 0.47 pm respectively from astrophysical observations of SN1987A and pulsars.[12,13] This example illustrates the connection and competition of precision laboratory experiments and astrophysical observations to cosmology.

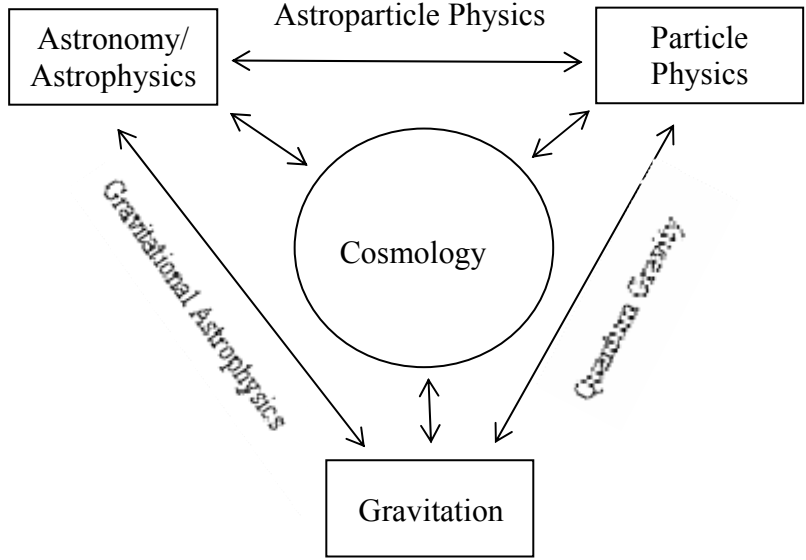

Fig. 1. The relation of cosmology to astronomy/astrophysics, particle physics, and gravitation.

Einstein Equivalence Principle (EEP) is the foundation of metric theories of gravity which constitute the gravitational basis of cosmology. EEP states that in all locality of spacetime, the local nongravitational physics is that of special relativity. This gives universal standards and serves as a theoretical basis of precision experiments. In precision experiments, any deviation and anomaly from this basis will be a deviation from EEP and will be important for cosmology. In this paper, we will first examine EEP and review the empirical foundation of metric theories of gravity (Section 2).

For studying the relations among equivalence principles, we used the following interaction Lagrangian density:

$$L_I = - (1/(16\pi))\chi^{ijkl} F_{ij} F_{kl} - A_k j^k (-g)^{(1/2)} - \Sigma_I m_I (ds_I)/(dt) \delta(\mathbf{x}-\mathbf{x}_I), \qquad (1)$$

where $\chi^{ijkl} = \chi^{klij} = -\chi^{klji}$ is the gravitational constitutive tensor of the gravitational fields (e.g., metric $g_{ij}$, (pseudo)scalar field $\varphi$, etc.) and $j^k$, $F_{ij} \equiv A_{j,i} - A_{i,j}$ have the usual meaning for electromagnetism.[14-18] Imposing Galileo Equivalence Principle, we found that the gravitational constitute tensor $\chi^{ijkl}$ must have the following form:

$$\chi^{ijkl} = (-g)^{1/2} [(1/2) g^{ik} g^{jl} - (1/2) g^{il} g^{kj} + \varphi\, e^{ijkl}], \qquad (2)$$

where $\varphi$ is a scalar or pseudoscalar function of the gravitational field and $e^{ijkl}$ is the completely antisymmetric symbol with $e^{0123} = 1$.[14-16] Since $\varphi$ is a scalar function, any constant factor in front of it could be absorbed.

The Lagrangian given by the last term in (2), i.e.,

$$L_I = -(1/(16\pi))\, \varphi\, e^{ijkl} F_{ij} F_{kl}, \qquad (3)$$

gives a pseudoscalar-photon interaction. Modulo a divergence, (3) is equivalent to [14)-16)]

$$L_I = (1/(8\pi))\, \varphi_{,i}\, e^{ijkl} A_j F_{kl}, \qquad \text{(mod div)}. \qquad (4)$$

The special case $\varphi_{,i}$ = constant = $V_i$ is considered by Carroll, Field and Jackiw.[19), 20)]

Pseudoscalar-photon interaction (4) would induce a rotation of linear polarization of electromagnetic wave propagating with cosmological/astrophysical distance.[14), 19), 20)] For right circularly polarized electromagnetic wave, the propagation from a point $P_1 = \{x_{(1)}^i\} = \{x_{(1)}^0; x_{(1)}^\mu\} = \{x_{(1)}^0, x_{(1)}^1, x_{(1)}^2, x_{(1)}^3\}$ to point $P_2 = \{x_{(2)}^i\} = \{x_{(2)}^0; x_{(2)}^\mu\} = \{x_{(2)}^0, x_{(2)}^1, x_{(2)}^2, x_{(2)}^3\}$ will add a phase of $\alpha = \varphi(P_2) - \varphi(P_1)$ to the wave; for left circularly polarized light, the added phase will be opposite in sign.[14)] Linearly polarized electromagnetic wave is a superposition of circularly polarized electromagnetic waves. Its polarization vector will then rotate by an angle $\alpha$. Locally, the polarization rotation angle can be approximated by

$$\alpha = \varphi(P_2) - \varphi(P_1) = {}_i\Sigma_0^3 [\varphi_{,i} \times (x_{(2)}^i - x_{(1)}^i)] = {}_i\Sigma_0^3 [\varphi_{,i}\Delta x^i] = \varphi_{,0}\Delta x^0 + [{}_\mu\Sigma_1^3 \varphi_{,\mu}\Delta x^\mu]$$

$$= {}_i\Sigma_0^3 [V_i \Delta x^i] = V_0 \Delta x^0 + [{}_\mu\Sigma_1^3 V_\mu \Delta x^\mu]. \qquad (5)$$

The rotation angle in (5) consists of 2 parts -- $\varphi_{,0}\Delta x^0$ and $[{}_\mu\Sigma_1^3 \varphi_{,\mu}\Delta x^\mu]$. For light in a local frame, $|\Delta x^\mu| = |\Delta x^0|$. In Fig. 2, space part of the rotation angle is shown. The amplitude of the space part depends on the direction of the propagation with the tip of magnitude on upper/lower sphere of diameter $|\Delta x^\mu| \times |\varphi_{,\mu}|$. The time part is equal to $\Delta x^0 \varphi_{,0}$. ($\nabla \varphi \equiv [\varphi_{,\mu}]$)

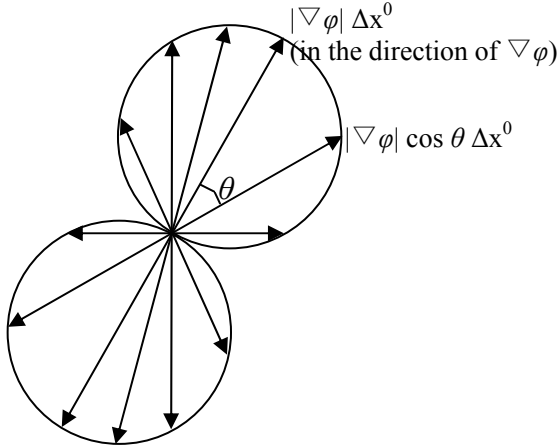

Fig. 2. Space contribution to the local polarization rotation angle -- $[{}_\mu\Sigma_1^3 \varphi_{,\mu}\Delta x^\mu] = |\nabla \varphi| \cos\theta\, \Delta x^0$. The time contribution is $\varphi_{,0} \Delta x^0$. The total contribution is $(|\nabla \varphi| \cos\theta + \varphi_{,0}) \Delta x^0$. ($\Delta x^0 > 0$)

In 2002, DASI[21)] (Degree Angular Scale Interferometer) successfully observed the polarization of the cosmological microwave background radiation. In 2003, WMAP[22)] observed the correlation of polarization with temperature anisotropy in the cosmological microwave background. From the polarization-temperature correlation in WMAP observation, we put a limit of 0.1 rad on the rotation of linear polarization of cosmological microwave background (CMB) propagation.[23),24)] Pseudoscalar-photon interaction is proportional to the gradient of the pseudoscalar field. From phenomenological point of view, this gradient could be neutrino number asymmetry current,[25), 26), 27)] other density current,[28)] or a constant vector.[19), 20)] In these situations, Lorentz invariance or CPT may[28)] or may not[25), 26), 27)] be effectively violated. Test of parity violation using CMB polarization observation was proposed in 1999.[29)] In this paper, we will discuss various situations. Better accuracy in CMB polarization observation is expected from PLANCK mission to be launched next year. A dedicated

CMB polarization observer in the future would probe this fundamental issue more deeply. The current constraints on the rotation angle are shown in Table I.

Table I. Constraints on cosmic polarization rotation from CMB (cosmic microwave background).

| Reference | Constraint [mrad] | Source data |
|---|---|---|
| Ni[23),24)] | ±100 | WMAP1[22)] |
| Feng, Li, Xia, Chen, and Zhang[30)] | -105 ± 70 | WMAP3[31)] & BOOMERANG (B03)[32)] |
| Liu, Lee, Ng[33)] | ±24 | BOOMERANG (B03)[32)] |
| Kostelecky and Mews[34)] | 209 ± 122 | BOOMERANG (B03)[32)] |
| Cabella, Natoli and Silk[35)] | -43 ± 52 | WMAP3[31)] |
| Xia, Li, Wang, and Zhang[36)] | -108 ± 67 | WMAP3[31)] & BOOMERANG (B03)[32)] |

In section 2, we review the $\chi$-g framework for analyzing equivalence principles and present the empirical foundation of EEP. In section 3, we discuss a few concepts related to polarization physics and present cosmic rotation of electromagnetic polarization. In section 4, we discuss CMB experiments and their constraints on the cosmic rotation of electromagnetic polarization. In section 5, we review possible causes of cosmic polarization rotation including neutrino number asymmetry, quintessence, Lorentz invariance violation and CPT violation. In section 6, we present an outlook.

## §2. Precision measurement and empirical foundation of relativistic gravity

The foundation of relativistic gravity at present rests on the equivalence of local physics to special relativity everywhere in spacetime. This equivalence is called the Einstein equivalence principle (EEP). Its validity guarantees the universal implementation of metrology and standards. Precision metrology/measurement, in turn, tests the validity of EEP.

The most tested part of equivalence is the Galileo equivalence principle (the universality of free-all). In the study of the theoretical relations between the Galileo equivalence principle and the Einstein equivalence principle, we use the $\chi$-g framework[15),16)] summarized in the interaction Lagrangian density given by (1). The gravitational constitutive tensor density $\chi^{ijkl}$ dictates the behavior of electromagnetism in a gravitational field and has 21 independent components in general. For a metric theory (when EEP holds), $\chi^{ijkl}$ is determined completely by the metric $g^{ij}$ and equals $(-g)^{1/2}[(1/2) g^{ik} g^{jl} - (1/2) g^{il} g^{kj}]$. Here we review the use this framework to look into the foundation of relativistic gravity empirically.

*No Birefringence.* The condition for no birefringence (no splitting, no retardation) for electromagnetic wave propagation in all directions in the weak field limit gives ten constraint equations on the $\chi$'s. With these ten constraints, $\chi$ can be written in the following form

$$\chi^{ijkl} = (-H)^{1/2}[(1/2)H^{ik} H^{jl} - (1/2)H^{il} H^{kj}]\psi + \varphi e^{ijkl}, \qquad (6)$$

where $H = \det(H_{ij})$ is a metric which generates the light cone for electromagnetic propagation.[17), 18), 37)] Note that (6) has the same form as (2) with $g^{ij}$ replaced by $H^{ij}$ and an added factor, 'dilation', $\psi$. Recently, Lämmerzahl and Hehl[38)] have shown that this non-birefringence guarantees, without approximation, Riemannian light cone, i.e., Eq. (6) holds without the assumption of weak field. To recover EEP, we need (i) no birefringence, (ii) no extra physical metric, (iii) no $\psi$ ('dilaton'), and (iv) no $\varphi$ (axion).

Eq. (6) is verified empirically to high accuracy from pulsar observations and from polarization measurements of extragalactic radio sources. With the null-birefringence observations of pulsar pulses and micropulses before 1980, the relations (6) for testing EEP are empirically verified to $10^{-14} - 10^{-16}$.[17),18),37)] With the present pulsar observations, these limits would be improved; a detailed such analysis is given by Huang.[39)] Analyzing the data from polarization measurements of extragalactic radio sources, Haugan and Kauffmann[40)] inferred that the resolution for null-birefringence is 0.02 cycle at 5 GHz. This corresponds to a time resolution of $4 \times 10^{-12}$ s and gives much better constraints. With a detailed analysis and more extragalactic radio observations, (6) would be tested down to $10^{-28}$-$10^{-29}$ at cosmological distances. In 2002, Kostelecky and Mews[41)] used polarization measurements of light from cosmologically distant astrophysical sources to yield stringent constraints down to $2 \times 10^{-32}$. The electromagnetic propagation in Moffat's nonsymmetric gravitational theory fits the $\chi$-$g$ framework. Krisher,[42)] and Haugan and Kauffmann[40)] have used the pulsar data and extragalactic radio observations to constrain it. It is interesting to note that just as the $\chi$-$g$ framework[15),16)] was being developed, an upper limit was set for polarization effects on gravitational deflection for radio waves passing through Sun's gravitational field.[43)] From now on, we assume (6).

*One Physical Metric and no 'Dilation' ($\psi$).* Let us now look into the empirical constraints for $H^{ij}$ and $\psi$. In Eq. (1), $ds$ is the line element determined from the metric $g_{ij}$. From Eq. (6), the gravitational coupling to electromagnetism is determined by the metric $H_{ij}$ and two (pseudo)scalar fields $\varphi$ 'axion' and $\psi$ 'dilaton'. If $H_{ij}$ is not proportional to $g_{ij}$, then the hyperfine levels of the lithium atom, the beryllium atom, the mercury atom and other atoms will have additional shifts. But this is not observed to high accuracy in Hughes-Drever experiments.[44)] Therefore $H_{ij}$ is proportional to $g_{ij}$ to certain accuracy. Since a change of $H^{ik}$ to $\lambda H^{ij}$ does not affect $\chi^{ijkl}$ in Eq. (6), we can define $H_{11} = g_{11}$ to remove this scale freedom.[17),18),45)]

In Hughes-Drever experiments,[44)] $\Delta m/m \leq 0.5 \times 10^{-28}$ or $\Delta m/m_{e.m.} \leq 0.3 \times 10^{-24}$ where $m_{e.m.}$ is the electromagnetic binding energy. Using Eq. (6) in Eq. (1), we have three kinds of contributions to $\Delta m/m_{e.m.}$. These three kinds are of the order of (i) ($H_{\mu\nu} - g_{\mu\nu}$), (ii) ($H_{0\mu} - g_{0\mu}$)v, and (iii) ($H_{00} - g_{00}$)v$^2$ respectively.[17),18),45)] Here the Greek indices $\mu$, $\nu$ denote space indices. Considering the motion of laboratories from earth rotation, in the solar system and in our galaxy, we can set limits on various components of ($H_{ij} - g_{ij}$) from Hughes-Drever experiments as follows:

$$| H_{\mu\nu} - g_{\mu\nu} | / U \leq 10^{-18},$$
$$| H_{0\mu} - g_{0\mu} | / U \leq 10^{-13} - 10^{-14},$$
$$| H_{00} - g_{00} | / U \leq 10^{-10}, \qquad (7)$$

where $U$ ($\sim 10^{-6}$) is the galactic gravitational potential.

Eötvös-Dicke experiments are performed on unpolarized test bodies.[46)] In essence, these experiments show that unpolarized electric and magnetic energies follow the same trajectories as other forms of energy to a certain accuracy. The constraints on Eq. (6) are

$$| 1-\psi | / U < 10^{-10} \qquad (8)$$

and

$$| H_{00} - g_{00} | / U < 10^{-6} \qquad (9)$$

where $U$ is the solar gravitational potential at the earth.

In 1976, Vessot and Levine[47)] used an atomic hydrogen maser clock in a space probe to test and confirm the metric gravitational redshift to an accuracy of $1.4 \times 10^{-4}$, i. e.,

$$| H_{00} - g_{00} | / U \leq 1.4 \times 10^{-4}, \qquad (10)$$

where $U$ is the change of earth gravitational field that the maser clock experienced.

The constraint (8) on the dilaton $\psi$ is stringent. However, with an appropriate mass or potential, the interaction range for dilaton[48),49)] or chameleon[50)] becomes intermediate and the associated constraint (8) becomes mild because the corresponding interaction becomes smaller.

With (i) no birefringence, (ii) no extra physical metric, (iii) no $\psi$ ('dilaton'), we arrive at the theory (1) with $\chi^{ijkl}$ given by (2), i. e., an axion theory.[51)] The current constraints on $\varphi$ from CMB polarization observations are listed in Table I. (See also Section 4.)

## §3. Rotation of polarization

For the theory (1) with the gravitational constitutive tensor (2), the electromagnetic wave propagation equation is

$$F^{ik}{}_{,k} + e^{ikml} F_{km} \varphi_{,l} = 0, \qquad (11)$$

in a local inertial (Lorentz) frame of the $g$-metric. Analyzing the wave into Fourier components, imposing the radiation gauge condition, and solving the dispersion eigenvalue problem, we obtain $k = \omega + (n^\mu \varphi_{,\mu} + \varphi_{,0})$ for right circularly polarized wave and $k = \omega - (n^\mu \varphi_{,\mu} + \varphi_{,0})$ for left circularly polarized wave in the eikonal approximation.[14)] Here $n^\mu$ is the unit 3-vector in the propagation direction. The group velocity is

$$v_g = \partial\omega/\partial k = 1, \qquad (12)$$

independent of polarization. There is no birefringence. For linearly polarized wave, there is an induced rotation of polarization with an angle of $(n^\mu \varphi_{,\mu} + \varphi_{,0}) = \Delta\varphi = \varphi_2 - \varphi_1$ where $\varphi_1$ and $\varphi_2$ are the values of the scalar field at the beginning and end of the wave. When we integrate along light (wave) trajectory, the total polarization rotation (relative to no $\varphi$-interaction) is $\Delta\varphi = \varphi_2 - \varphi_1$ where $\varphi_1$ and $\varphi_2$ are the values of the scalar field at the beginning and end of the wave. When the propagation distance is over a large part of our observed universe, we call this phenomenon cosmic polarization rotation.

Here we must say something about nomenclature.

Birefringence, also called double refraction, refers to the two different directions of propagation that a given incident ray can take in a medium, depending on the direction of polarization. The index of refraction depends on the direction of polarization.

Dichroic materials have the property that their absorption constant varies with polarization. When polarized light goes through dichroic material, its polarization is rotated due to difference in absorption in two principal directions of the material for the two polarization components. This phenomenon or property of the medium is called dichroism.

In a medium with optical activity, the direction of a linearly polarized beam will rotate as it propagates through the medium. A medium subjected to magnetic field becomes optically active and the associated polarization rotation is called Faraday rotation.

Cosmic polarization rotation is neither dichroism nor birefringence. It is more like optical activity, with the rotation angle independent of wavelength. Conforming to the common usage in optics, one should not call it cosmic birefringence.

## §4. CMB constraints on the cosmic polarization rotation $\Delta\varphi$

In this section, we review and compile the constraints of various analyses from CMB polarization observations. Ten years ago, Nodland and Ralston announced in a paper[52)] that they found an additional rotation of synchrotron radiation from distant radio galaxies and quasars which is independent of wavelength. However, other people did not find this in their analysis and put a limit of $\Delta\varphi \leq 0.17\text{-}1$ over cosmological distance from polarization observations of radio galaxies.[53)-58)] In particular, Cimanti, di Serego Alighieri, Field, and Fosbury had found no rotation within 10 degrees (0.17 rad) for the optical/UV polarization of radio galaxies for all radio galaxies with 0.5<z<2.6 in their list.[53),54)] There is also no rotation within 10 degrees for a recent update with z>2.0 (up to z =4.1).[59)]

In 2002, DASI microwave interferometer observed the polarization of the cosmic background.[21)] E-mode polarization is detected with 4.9 σ. The TE correlation of the temperature and E-mode polarization is detected at 95% confidence. This correlation is expected from the Raleigh scattering of radiation. However, with the (pseudo)scalar-photon interaction (3)/(4), the polarization anisotropy is shifted differently in different directions relative to the temperature anisotropy due to propagation; the correlation will then be downgraded. In 2003, from the first-year data (WMAP1), WMAP found that the polarization and temperature are correlated to more than 10 σ.[22)] This gives a constraint of about $10^{-1}$ for $\Delta\varphi$.[23),24)]

Further results of CMB polarization observations came out after 2003. CBI (Cosmic Background Imager) reported their observation of E-mode polarization with 8.9 σ.[60)] BOOMERANG detected a TE correlation with a statistical significance >3.5 σ, using their 2003 flight data (BOOMERANG03).[32)] From three-year data, WMAP reported E-mode polarization with a statistical significance >3 σ, and better TE correlation.[31)] These experiments have yielded polarization results of a wide range of angles.

In 2006, Feng et al. used BOOMERANG03 and WMAP3 data for fitting polarization rotation and found $\Delta\varphi = -6 \pm 4$ deg at 1 σ confidence level.[30)] They interpreted this as possibly be resulted from the CPT and Lorentz violations. Including the information of TB and EB power spectra, Xia et al. found the polarization rotation $\Delta\varphi = -6.2 \pm 3.8$ deg at 1 σ confidence level with the same interpretation.[36)] Liu, Lee and Ng in their work estimated that $\Delta\varphi = \pm 1.4$ deg at 1 σ confidence level.[33)] Kostelecky and Mews used the results from the BOOMERANG03 experiment and gave a cosmic polarization rotation $\Delta\varphi = 12 \pm 7$ deg.[34)] Cabella, Natoli and Silk performed a wavelet analysis of the WMAP03 data in search for a parity violating signal and set a 1 σ limit on the cosmic polarization rotation at $\Delta\varphi = -2.5 \pm 3.0$ deg.[35)] These results are all compiled in Table I. Although these results look different at 1 σ level, they are all consistent with null detection and with one another at 2 σ level. We turn to the interpretation of cosmic polarization rotation in the next section.

## §5. Neutrino number asymmetry, quintessence, Lorentz Invariance and CPT

Geng, Ho and Ng proposed a new type of effective interactions in terms of the CPT-even dimension-six Chern-Simons-like term to generate the cosmic polarization rotation, and used the neutrino number asymmetry to induce a non-zero polarization rotation angle in the data of the CMB polarization.[25),26),27)] They found that the rotation effect can be of the order of magnitude of 0.01-0.1 rad or smaller.

The Lagrangian (3) can be extended to all gauge fields with (4) valid for the Abelian case, and the same implication on the relations of equivalence principles holds.[61)] Bamba, Geng and Ho considered hypercharge field together with baryon current and found that the hypermagnetic baryogenesis is experimentally viable.[62)]

In the work of Feng et al.,[30)] they proposed CPT violation and dynamical dark energy. In a more recent paper, Li et al.[63)] considered baryo/leptogenesis with cosmological CPT violation as a possible cause and gave a 1σ limit on their fermion current-curvature coupling parameter $\delta = -0.011 \pm 0.007$. Liu, Lee and Ng gave constraints on the coupling between the quintessence and the pseudoscalar of electromagnetism.[33)] Kostelecky and Mews extended their SME[64)] (Standard Model Extension) whose electromagnetic sector is the same as that in (1) with the gravitational constitutive tensor set to constant, to include some higher order terms, and gave constraints on various terms from BOOMERANG. The most precise constraint is on one of the SME parameter which gives cosmic polarization rotation $\Delta\varphi$. Their constraint is shown in Table I. Alfaro et al.[65)] critically examined the quantization of the Lorentz invariance violating electrodynamics and showed that it is consistent either for a light-like cosmic anisotropy axial-vector or for a time-like one, when in the presence of a bare photon mass.

### §6. Outlook

We started using the phenomenological Lagrangian (1) with a gravitational constitutive tensor. Astrophysical observations and precision laboratory experiments constrain the gravitational constitutive tensor $\chi^{ijkl}$ to (2) to high accuracy with one degree of freedom left. This opens up the (pseudo)scalar-photon interaction (3)/(4) [the axion theory for electromagnetism] which in turn predicts cosmic polarization rotation. Possible interpretations for a positive detection include (pseudo)scalar field cosmology,[66] inflation, quintessence, 'spontaneous polarization' in fundamental law of electromagnetic propagation, neutrino number asymmetry, other number asymmetry, Lorentz invariance violation, CPT violation, etc.

CMB polarization experiments give the best current constraints on the cosmic polarization rotation to about 0.1 (100 mrad). Planck Surveyor will be launched in 2008 with better polarization-temperature measurement and will give a sensitivity to $\Delta\varphi$ of $10^{-2}$-$10^{-3}$ (1-10 mrad).[23],[24] A dedicated future experiment on cosmic microwave background polarization will reach $10^{-5}$-$10^{-6}$ (1-10 μrad) $\Delta\varphi$-sensitivity.[23],[24] Xia et al.[67] recently simulated with accuracy of Planck and a dedicated CMB polarization experiment and give the standard deviation of the cosmic polarization rotation angle to be $\sigma = 10^{-3} = 1$ mrad for PLANCK and $\sigma = 4.5 \times 10^{-5} = 45$ μrad for CMBpol. This confirms our estimate largely. This is very significant as a positive result may indicate that our patch of inflationary universe has a 'spontaneous polarization' in fundamental law of electromagnetic propagation influenced by neighboring patches and we can 'observe' neighboring patches through a determination of this fundamental physical law; if a negative result turns out at this level, it may give a good constraint on superstring theories as axions are natural to superstring theories.[23],[24] It may also measures or gives constraint on neutrino number asymmetry, other number asymmetry, Lorentz invariance violation, CPT violation, etc.

### Acknowledgements

We would like to thank the National Natural Science Foundation of China (Grant Nos. 10475114 and 10778710) and the Foundation of Minor Planets of Purple Mountain Observatory for support.